# Nature of segregation of reactants in diffusion controlled A+B reactions: Role of mobility in forming compact clusters

Panos Argyrakis and Raoul Kopelman

Department of Chemistry, University of Michigan, Ann Arbor, Mi 48109 USA

Abstarct

We investigate the A+B=0 bimolecular chemical reaction taking place in low-dimensional spaces when the mobilities of the two reacting species are not equal. While the case of different reactant mobilities has been previously reported as not affecting the scaling of the reactant densities with time, but only the pre-exponential factor, the mechanism for this had not been explained before. By using Monte-Carlo simulations we show that the nature of segregation is very different when compared to the normal case of equal reactant mobilities. The clusters of the mobile species are statistically homogeneous and randomly distributed in space, but the clusters of the less mobile species are much more compact and restricted in space. Due to the asymmetric mobilities, the initial symmetric random density fluctuations in time turn into asymmetric density fluctuations. We explain this trend by calculating the correlation functions for the positions of particles for the several different cases.

## I. Introduction

Over the past thirty years [1-7] there has been considerable interest in chemical reactions occurring in the condensed phase, when the process in question can be characterized as reaction-diffusion limited. In such a system there is always a competition between these two mechanisms, namely the reaction of the two species coming in contact and the diffusion of these species as they move in space. The resulting overall behavior that one observes reflects a combination of these two factors. The interest in such processes stems from the fact that the rate of the reaction is highly non-classical. For a bimolecular reaction it turns out that the rate is strongly dependent on the dimensionality of the system, and scales with time, but highly non-linearly, with exponents that are always smaller than one (1). This was first suggested by Ovchinnikov and Zeldovich [5], was verified numerically by Toussaint and Wilczec [6], and thereafter discussed by a plethora of authors [see review paper, Ref. 7]. The startling observation was also made at that time that for the bimolecular reaction of two different species (say, A and B) there is a spontaneous self-segregation of these two types of reactants, into clusters of each kind, this being the reason for the slowing-down of the reaction: The bulk of the reactants are shielded inside the clusters formed, while the reaction now proceeds only along the front or periphery of the clusters, because A can only react with B and vice-versa, but not A with A and B with B.

There have been only a few experimental systems reported in the literature, up to now, that follow such a scheme of the A+B model. Recently [8], the kinetics of the surface reaction $NO + H^+ + NO_3^- \rightarrow NO_2 + HNO_2$ on BaNa-Y has been reported, which is a system in which one of the reactants is immobile. Particularly it was shown that $H^+$ is mobile, while $NO_3^-$ is immobile. This raises certainly the question whether the kinetics of such a reaction is dependent at all on the mobility of the reacting species.

In a previous work [9] we had dealt in detail with the problem of the mobilities of the two reactants, testing if the non-classical effects indeed hold when the mobilities of the reactants are different. We had examined only the case of a 1-dimensional reaction, in which only 1 of the 2 species is mobile while the second one is immobile. We have found indeed that while the absolute densities were different from the case when both species are mobile, the scaling of the A+B reaction is not affected at all. We explained this observation by looking at the nearest neighbor distance distribution function (NNDDs). However, no geometrical explanation was given at that time as to how the clusters are formed (shaped) as a result of the different mobilities. In the current paper we examine both the 1-dimensional and 2-dimensional case and we find indeed a very unique behavior for the shapes of the clusters formed. The purpose of this paper is to investigate and explain these shapes. The main effect that we report on is shown in Figure 1, where we present the same A+B reaction in 2-dimensional space for two cases. In the first case the two species have the same mobility, while in the second case the first species has a mobility of m=1 (meaning one transition to a nearest neighbor site in one arbitrary time unit), while the second species is totally immobile, m=0. After the reaction has taken place for a time long enough for segregation to appear, we observe the following: For case (a) we see the typical clustering, as has usually been observed by many authors [5-7]. For case (b) we see that the immobile species are closely “bunched” together, into compact clusters, in a much smaller area of the system. The segregation has occurred very clearly but now the A and B clusters are not symmetrical, as in the case where both species are mobile. The A clusters are extended in space, while the B clusters are more compact. Notice that in both cases the concentration of A and B are always the same at all times, even though one might mistakenly assume that in case (b) the A population is larger than the B population. In Section II we describe in detail the method of simulations. In Section III we calculate the position-position correlation function for the two different cases and explain this difference in behavior. Finally, in section IV we present a summary of the results.

## II. Method of simulations

The A+B reaction is simulated in the usual way as has been reported in the literature by several groups [4,6,7,9,10]. For simplicity we assume only nearest neighbor transitions. When all particles present have made one such nearest neighbor transition this constitutes one Monte-Carlo step (MCS). We treat here the case when the second species is totally immobile, meaning that the B species never move, their mobility being m=0. Intermediate values of m can also easily be investigated, but this is not expected to change the nature of the current findings. All A and B particles are distributed initially, at time t=0, at random, on a lattice of size L=1,000,000 (1-dimensional), and square lattice geometry of size 5000x5000 (2-dimensional), with initial particle density $\rho_0=\rho_A=\rho_B=0.4$. All particles perform a normal random walk. The A particles react with B (and the B with A) upon encounter with perfect efficiency. This happens when they occupy the same lattice site at the same time. During this reaction both reacting particles are removed from the system. The A particles cannot react with other As, and similarly for the Bs. Cyclic boundary conditions are used at the edge of the lattice. We use the excluded volume principle, in which only one particle may occupy a single lattice site during one time. We monitor the decay of the particle density as a function of time and the reaction rate as a function of the density.

## II. Results and discussion

Following the pictorial presentation in Fig. 1, we show in Figure 2 the dependence of the density of the reacting species on time. As has been widely considered in the past the scaling of the density with time goes as [7]:

$$\frac{1}{\rho} - \frac{1}{\rho_0} = kt^f$$

Thus, in Figure 2 we plot the quantity $1/\rho - 1/\rho_0$ vs. time in log-log axes in order to recover the exponent f. We show the cases of 1-dimensional and 2-dimensional reactions, for both models. We clearly see that the slopes in both cases are the same, meaning that the reaction rates scale the same and their time exponents do not depend on the mobility of the reactants. This is in agreement with our previous work [9,10]. The slope for the 1-dimensional case is 0.259, while it is 0.53 for the 2-dimensional case. As has been discussed in the past, there is clearly an early time effect where the slope is not steady at all, and also some finite-size effect at long times. Therefore, the choice of proper region for calculating the slope is between these two limits and contains some uncertainty about the best range where to calculate it. Hence, there is some deviation from the theoretically expected values of 0.25, and 0.500, respectively.

Figure 3 contains the position-position correlation function, C(r), for the cases of reactions studied in this work, for one dimensional lattices. This function is derived by calculating the population of all particles away from a given particle, as a function of distance. This is done successively for all particles present in the system and is averaged over many realizations. We use a space interval of $\Delta r$=10 lattice units and we go out up to 1000 lattice units away from the particle investigated. In part (a) we show the A-B function for the case where both species are mobile, for four different times during the course of the reaction. As the reaction progresses, we observe the expected result, i.e. that the asymptotic value of C(r,t)=1 is reached at longer and longer distances. In part (b) we show the correlation function for the same case (both species mobile) for the A-A and the B-B functions, summed together due to their symmetrical behavior. Again we notice the same observation as in the previous part, i.e. C(r,t) approaching 1 at longer times. In parts (c), (d), and (e) we show the case where A is mobile but B is immobile, in (c) for the A-A function, in (d) for the A-B function and in (e) for the B-B function. For all these cases we notice the same following observations: all curves are shifted to the left, compared to (a) and (b). This is a manifestation of the compact cluster effect described in Figure 1, i.e. the distances between the B particles are now much shorter due to the "bunching" effect shown in Figure 1, making the B clusters much more compact than the corresponding A clusters.

Analogous effects are shown in Figure 4, parts (a)-(f), for the reaction taking place on two dimensional lattices. The difference between the cases of (1) both species being mobile vs. (2) only one species being mobile is even more dramatic now than the one shown for the one dimensional case, e.g. when comparing 4(c) to 4(d). Again, this behavior is the result of an enhanced compactness of the immobile species clusters. This enhanced compactness may be due to the fact that in two dimensions the number of nearest neighbors is 4 (vs. 2 in one dimension), which gives more possibilities for the particles to move and react.

## IV. Summary

In this work we have examined some new effects shown by a bimolecular reaction in low-dimensional spaces, when only one of the two reactants is mobile, while the other one is immobile. The motivation to investigate this effect in detail has been the recent experimental work claiming to have attained such a mobile-immobile reaction. We performed simulations of such reactions in one dimensional and two dimensional lattices. As may be expected, we found the same type of scaling for both cases. However, there is a striking difference in the spatial distribution of the reactants for these two cases. While such distribution is homogeneous when the two reactants have equal mobilities, we showed that, when one species is immobile, its segregated clusters are not homogeneously distributed over the entire space as in the first case, but they are rather restricted to some small fraction of the available space. Apparently the inability to move leads to the annihilation of most isolated immobile species. The calculated correlation functions verify these conclusions. Interestingly, symmetric density fluctuations at time zero turn into asymmetric density fluctuations at longer times in this case of asymmetric mobilities. Finally, for reactions with finite but very unequal mobility of A and B, we expect analogous results at long but finite reaction times.

Acknowledgment: This work was supported by NSF/DMR  Grant No.0455330 (RK).

Figure captions

Figure 1: Schematic pictorial of a single realization of the 2-dimensional A+B reaction under different mobility conditions. In part (a) the reactants move in the conventional way, i.e. the mobilities of the A and B particles are identical. In part (b), we show exactly the same reaction (all parameters are identical) as in part (a), with the only difference being that particles B never move. Only the A particles move, but the reaction proceeds in the normal way. Notice the different type of clustering in the two plots. This plot is typical of the particle layout under these conditions, and several other realizations performed are qualitatively very similar.

Figure 2: Plot of the function $1/\rho - 1/\rho_0$ vs. time for the 1-dimensional and 2-dimensional A+B reactions. In each case we have the two lines, one for both species being mobile (top in each group of two lines), and one for one species being immobile (bottom line in each group of two lines). The results are averages of 20 runs on lattices of size 1 million (1-dimensional) and 5000x5000 (2-dimensional), with initial density $\rho_0$=0.40 for each of the species.

Figure 3: Correlation function C(r,t) vs. distance r for the 1-dimensional system. We give here all functions according to the nature of the reactants, i.e. the A-A and B-B function (like-like particles), and the A-B function (like-unlike), for both cases of mobilities, as follows: (a) both species mobile, A-B function, (b) both species mobile, the A-A and the B-B functions summed together, (c) A is mobile while B is immobile, the A-A function, (d) A is mobile while B is immobile, the A-B function, (e) A is mobile while B is immobile, the B-B function. The four curves correspond to the correlation function evaluated at four different times, t=100, 1000, 10000, and 100000 MCS (left to right).

Figure 4: Correlation function C(r,t) vs. distance r for the 2-dimensional system. We give here all functions according to the nature of the reactants, i.e. the A-A and B-B function (like-like particles), and the A-B function (like-unlike), for both cases of mobilities, as follows: (a) both species mobile, the A-A and the B-B functions summed together, (b) A is mobile while B is immobile, the A-A function and the B-B functions summed together, (c) both species mobile, the B-B function, (d) A is mobile while B is immobile, the B-B function, (e) both species are mobile, the A-B function, and (f) A is mobile while B is immobile, the B-B function. The four curves correspond to the correlation function evaluated at four different times, t=100, 1000, 10000, and 100000 MCS (left to right).

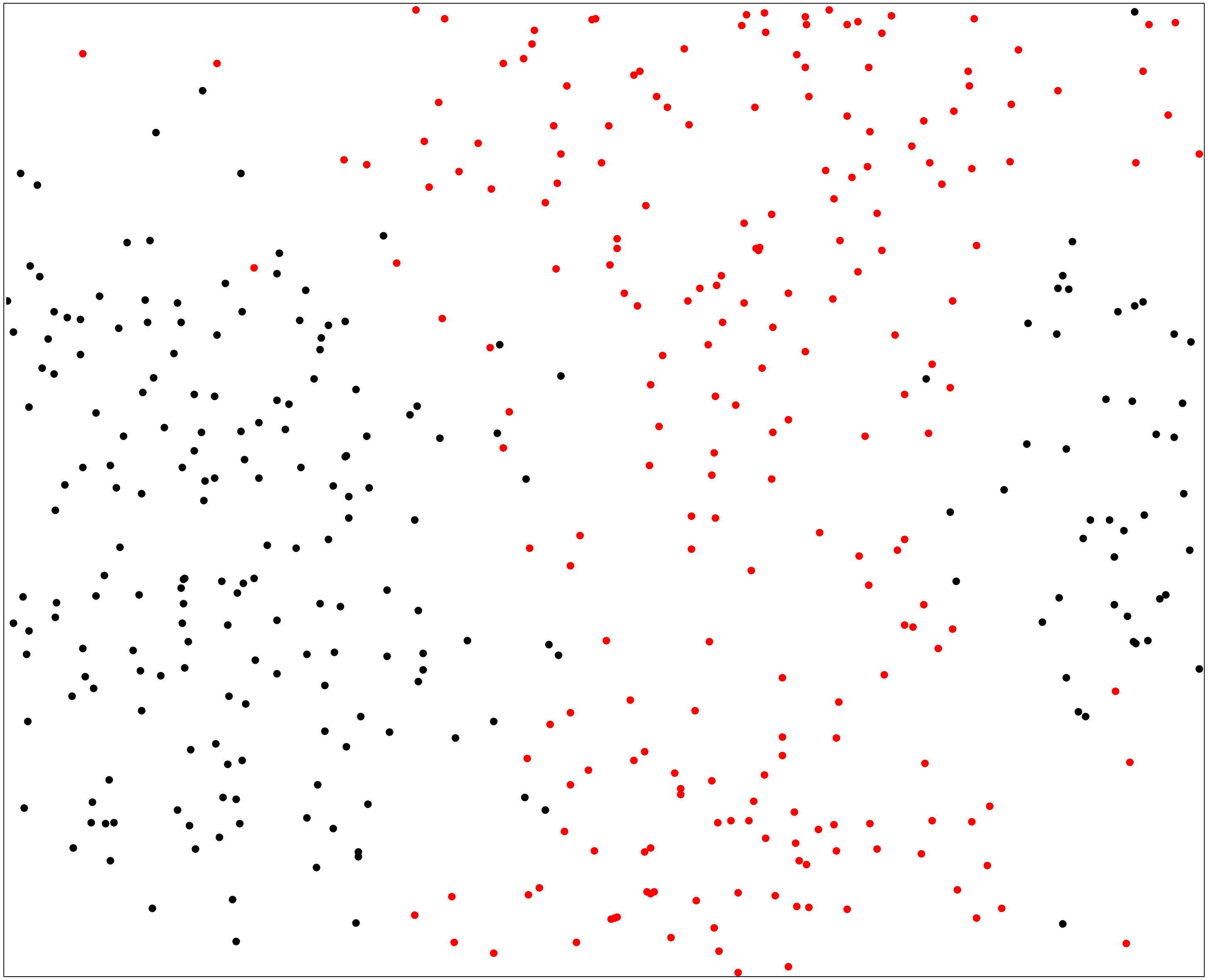

fig1a

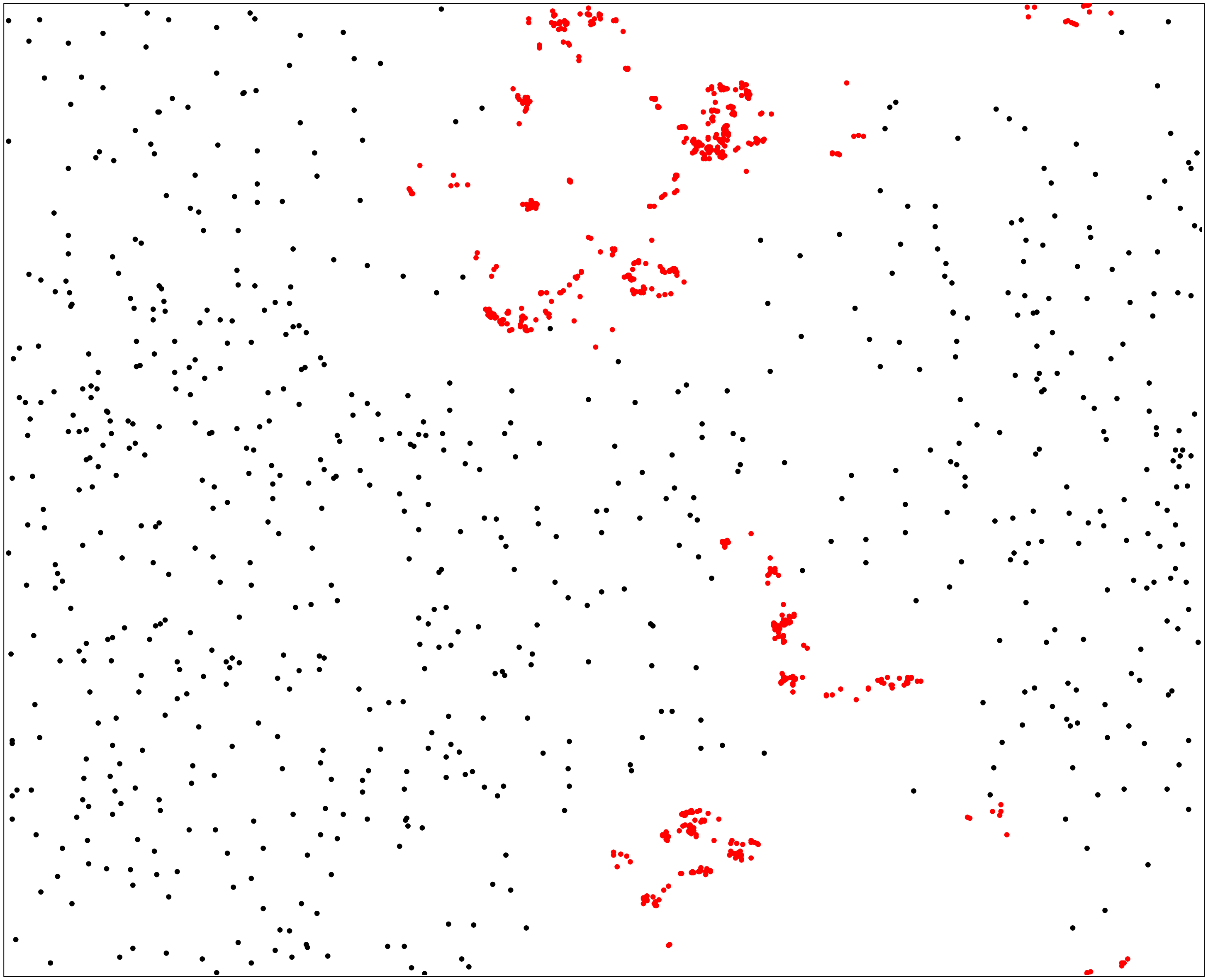

fig1b

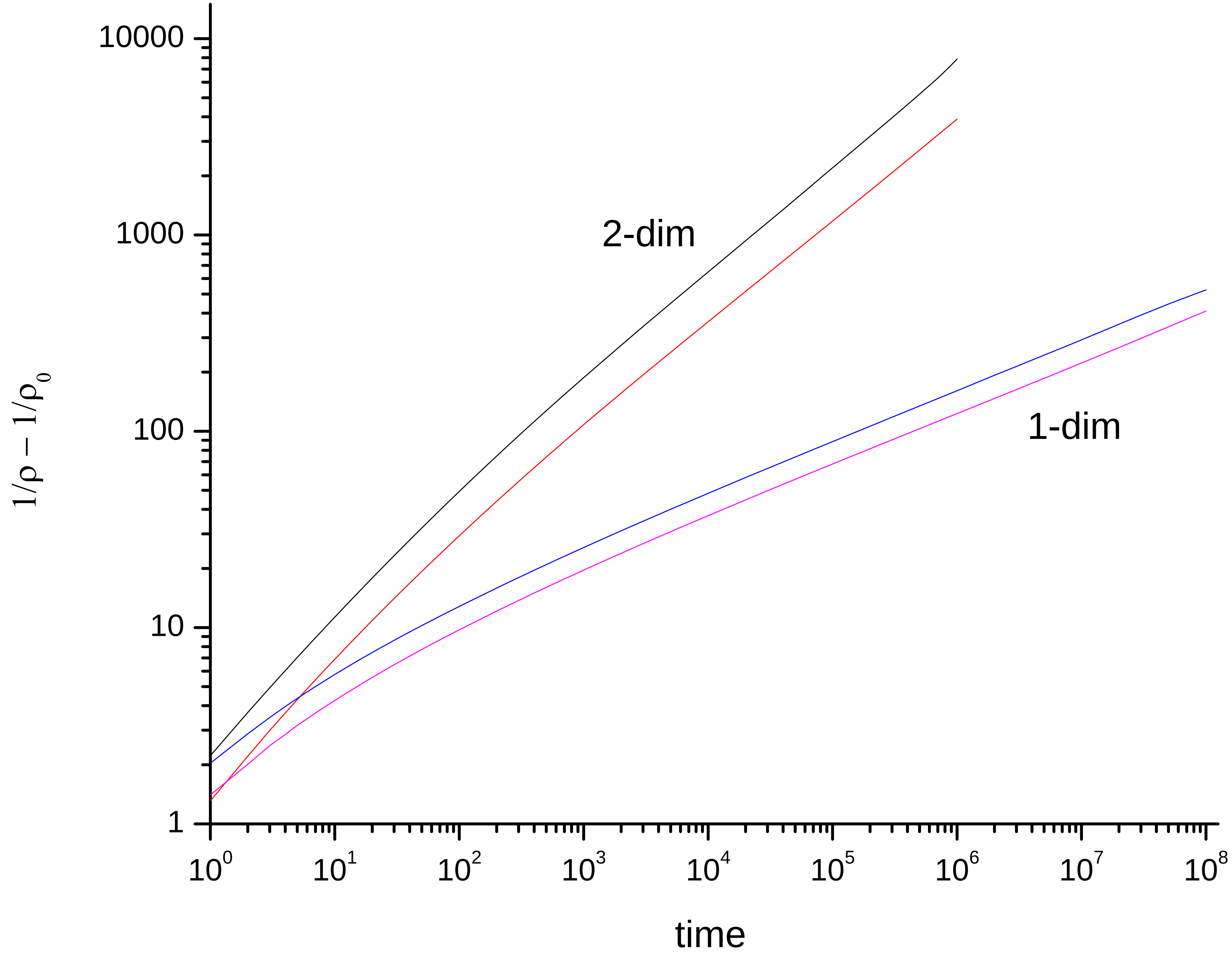

fig2

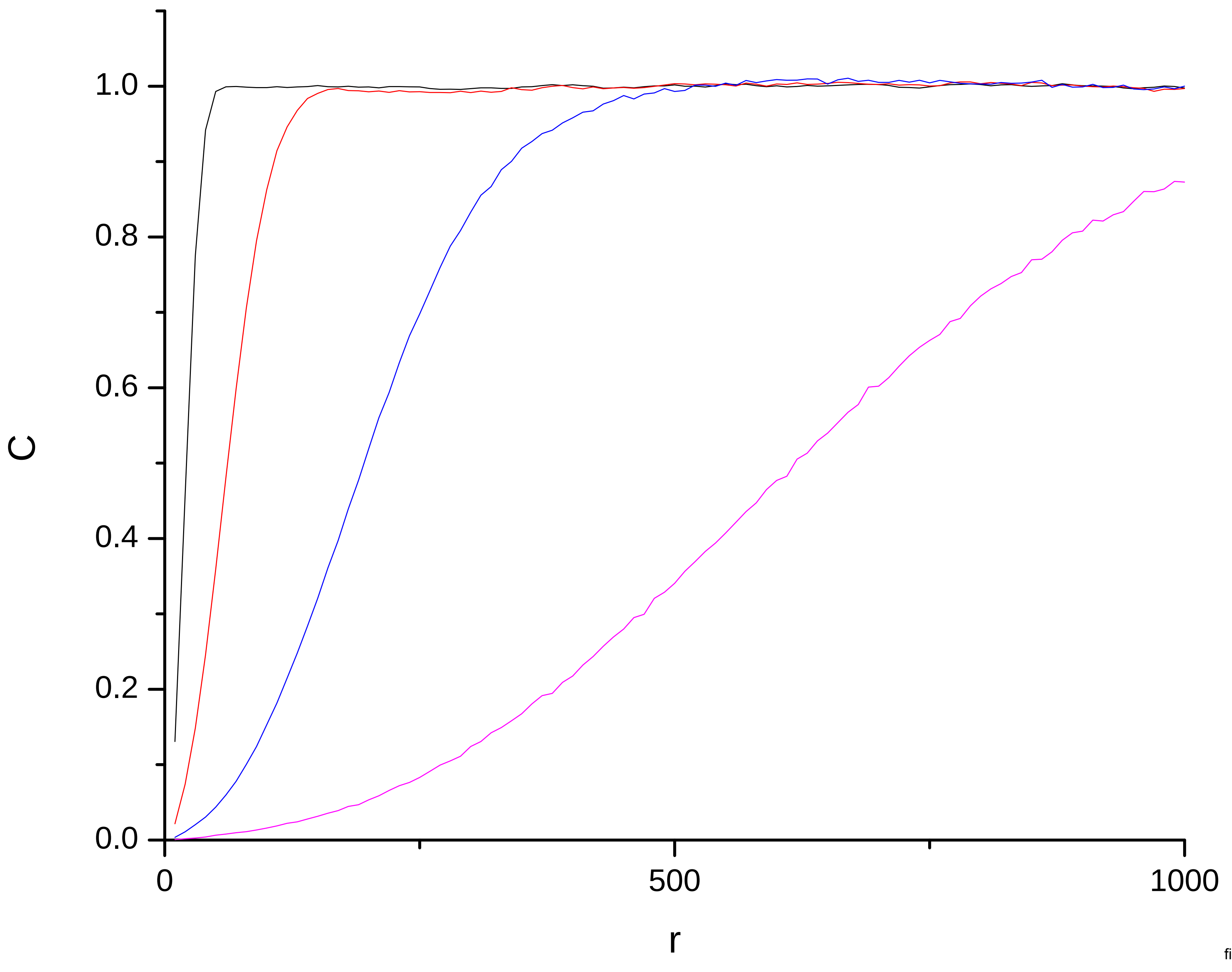

fig3a

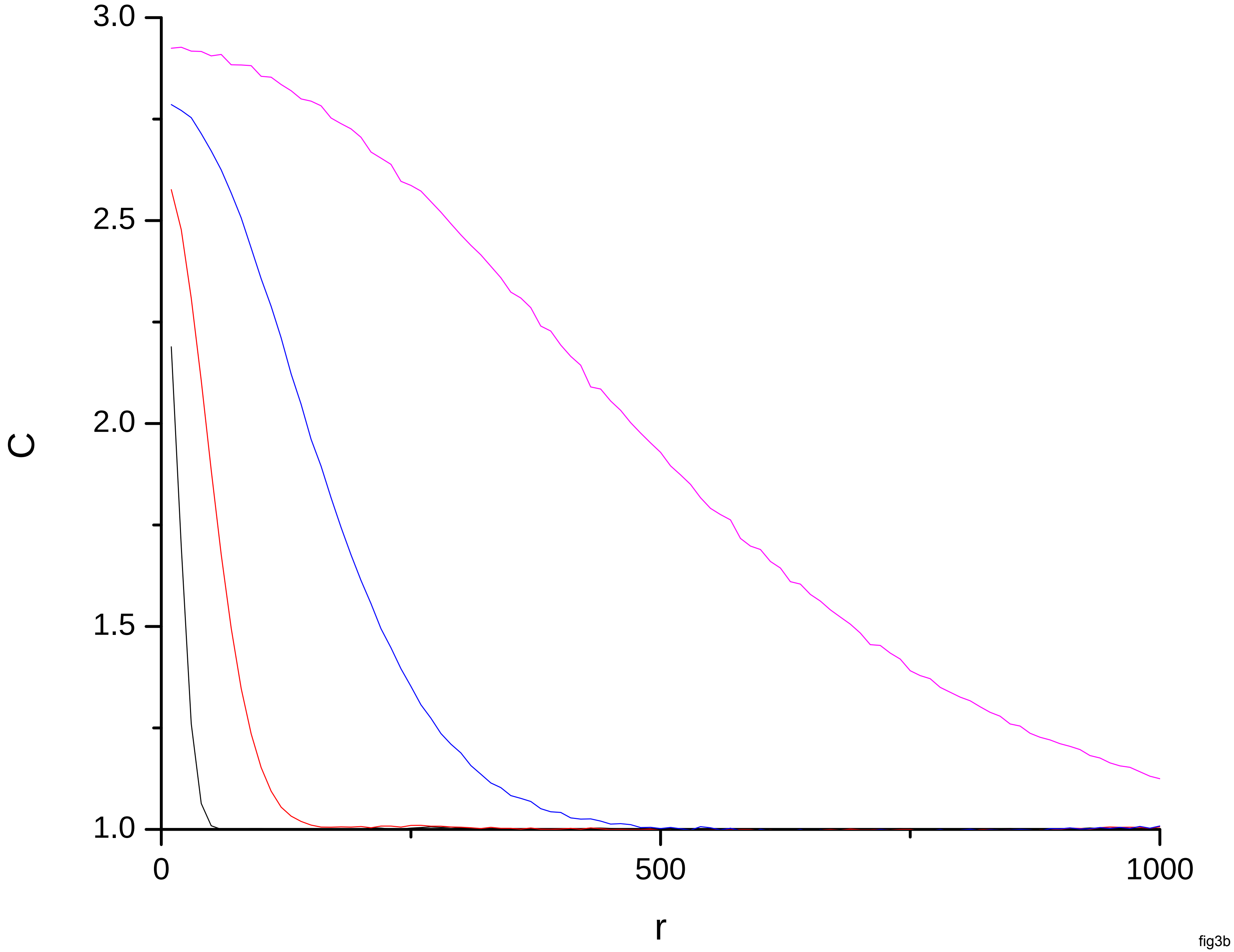

3.0
2.5
2.0
1.5
1.0
C
0
500
1000
r
fig3b

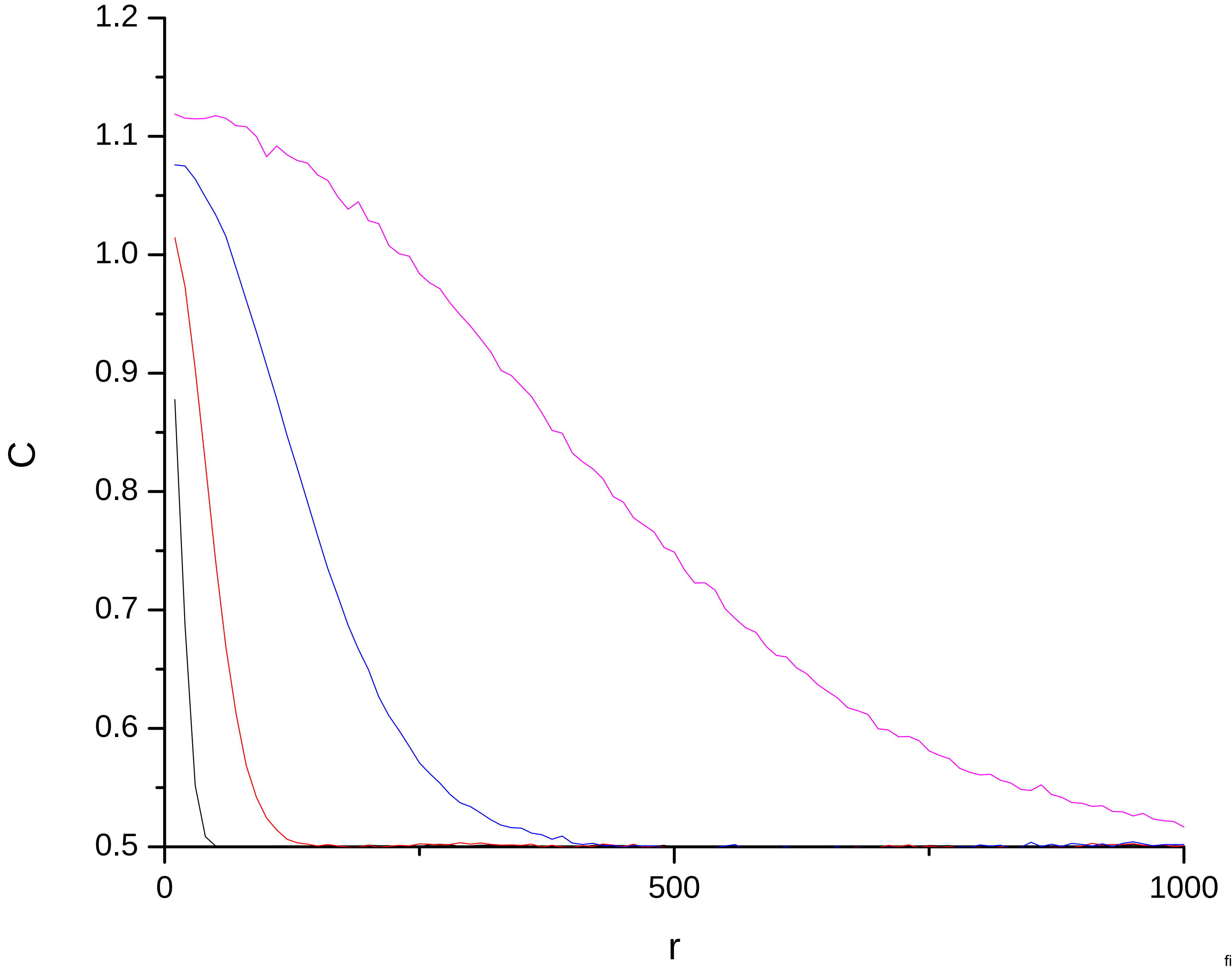

fig3c

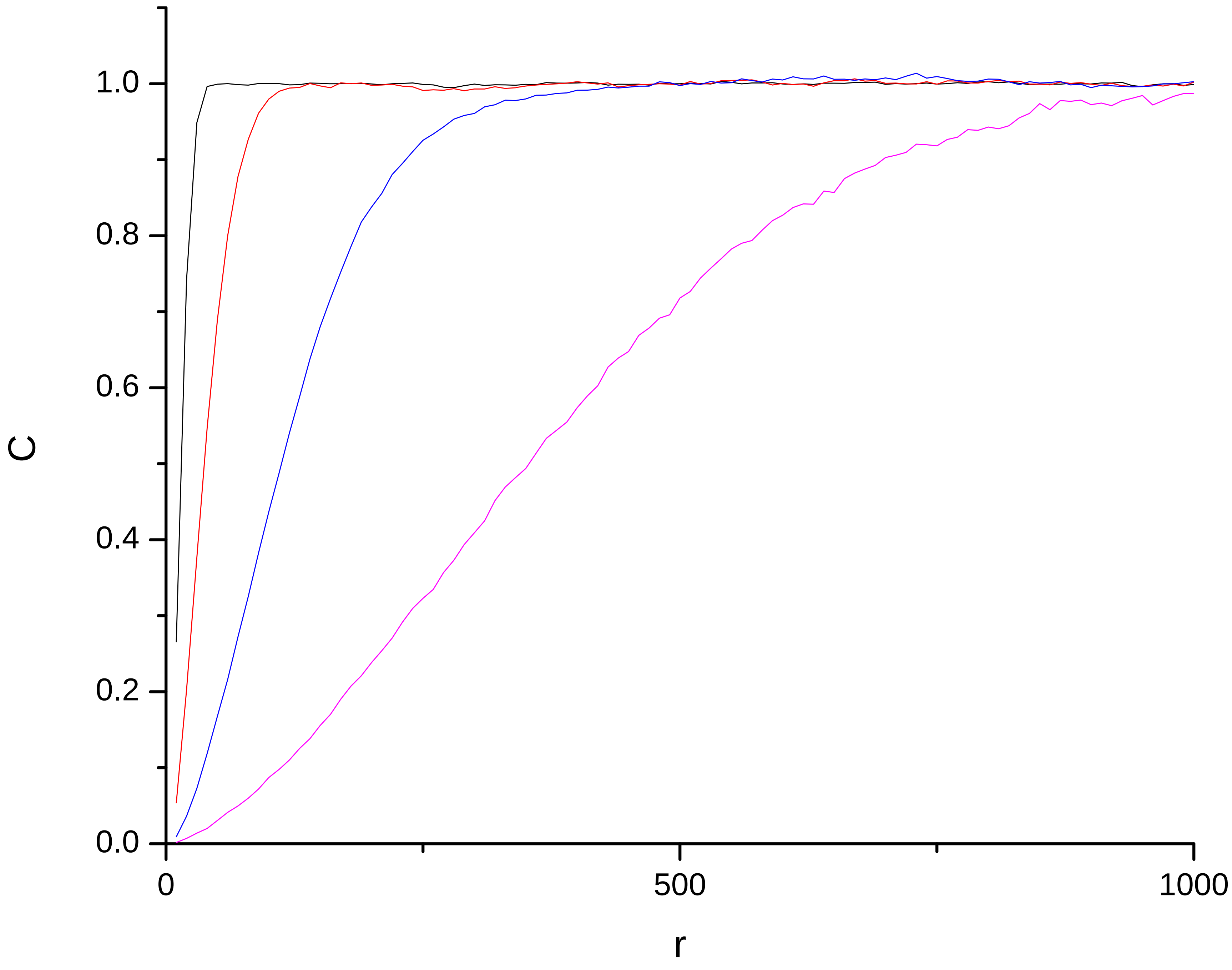

fig3d

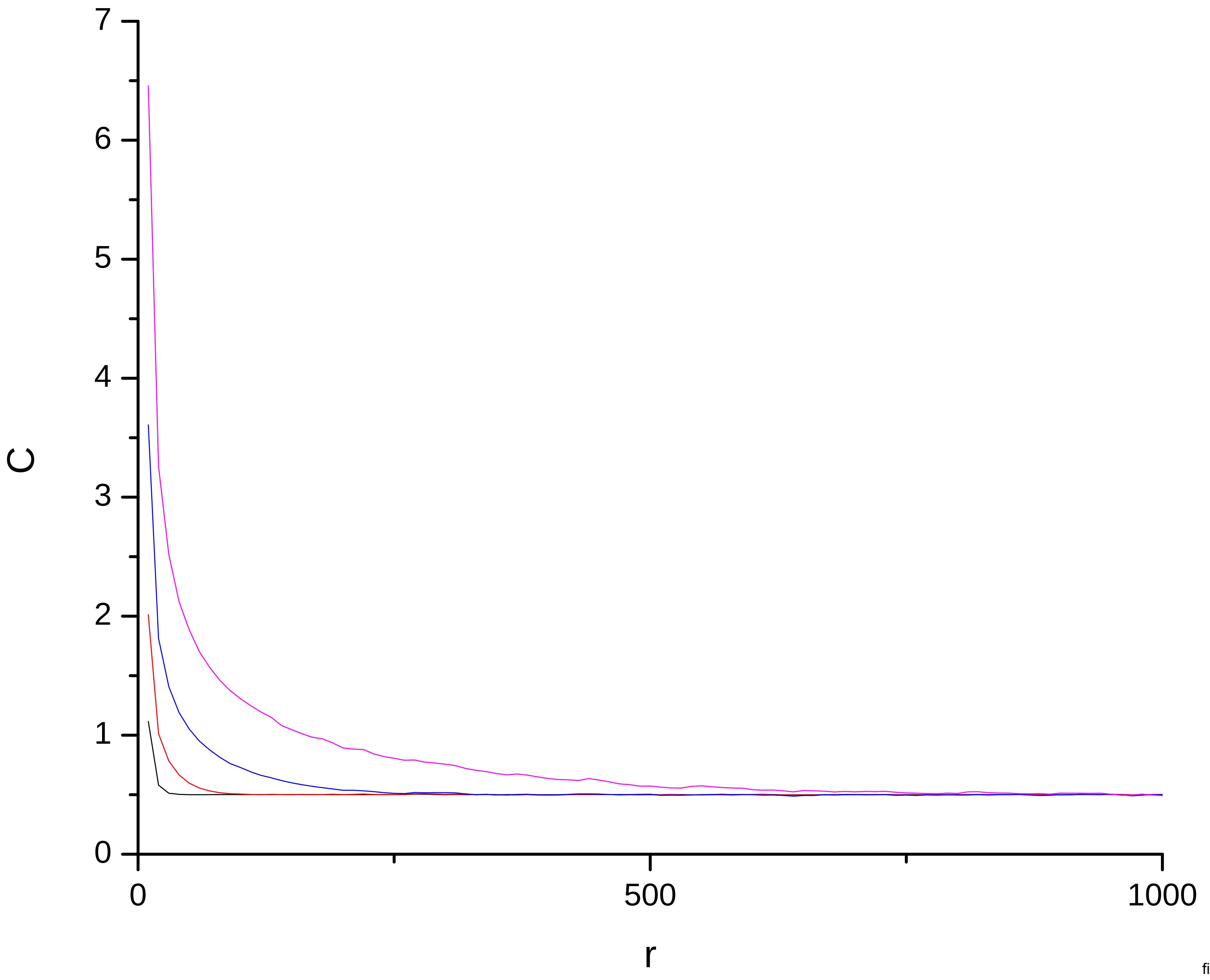

fig3e

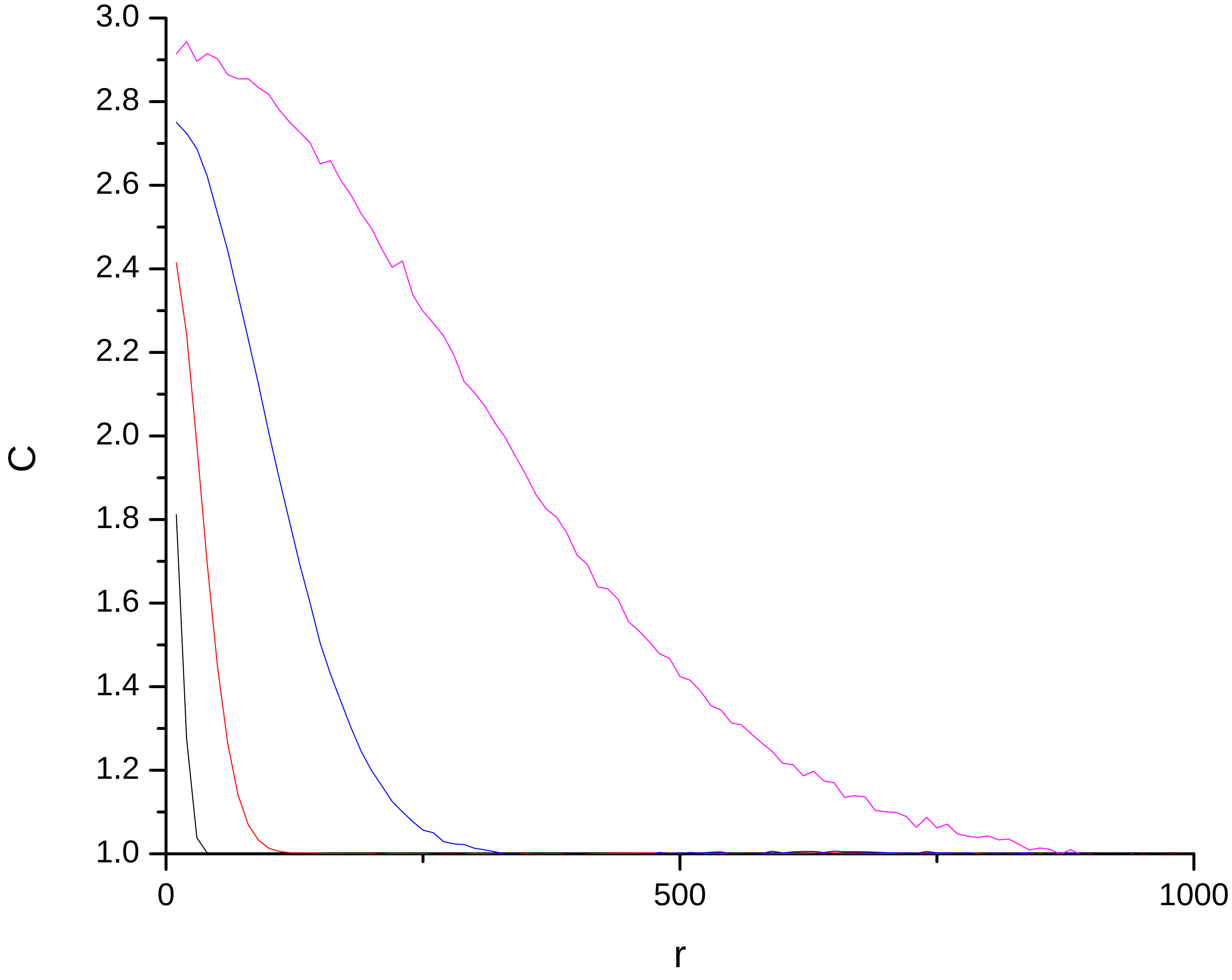

fig4a

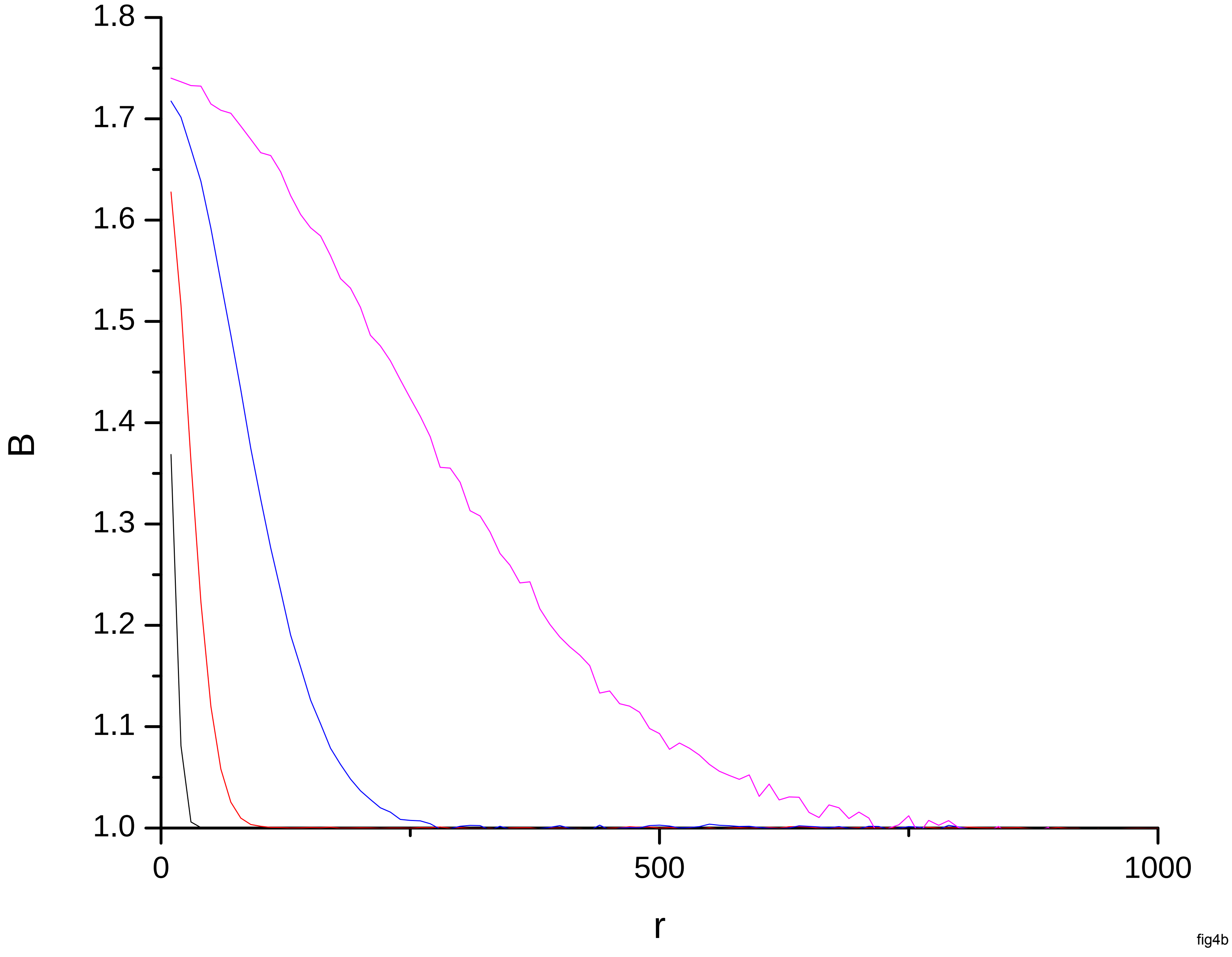
1.8
1.7
1.6
1.5
1.4
1.3
1.2
1.1
1.0
B
0
500
1000
r
fig4b

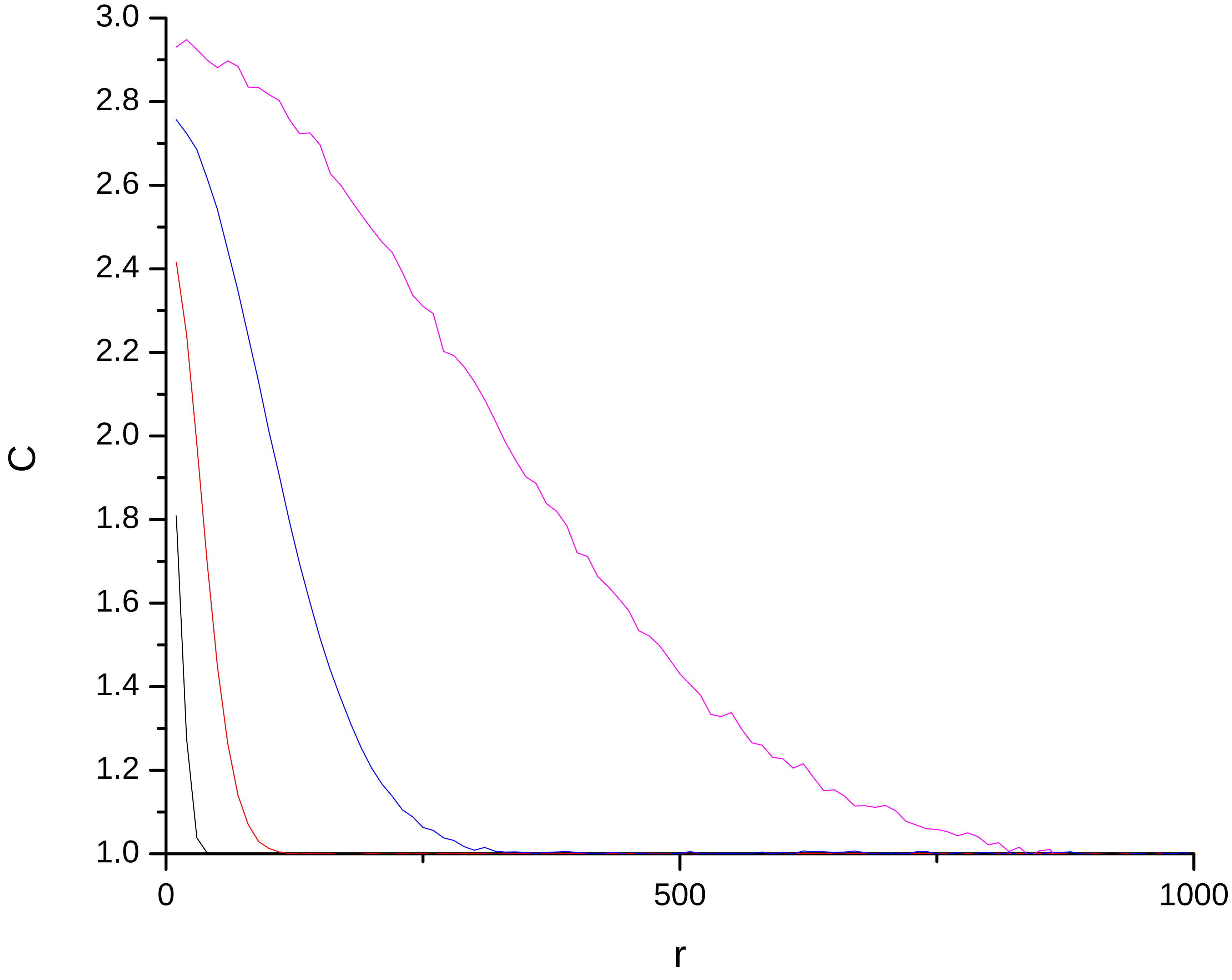

fig4c

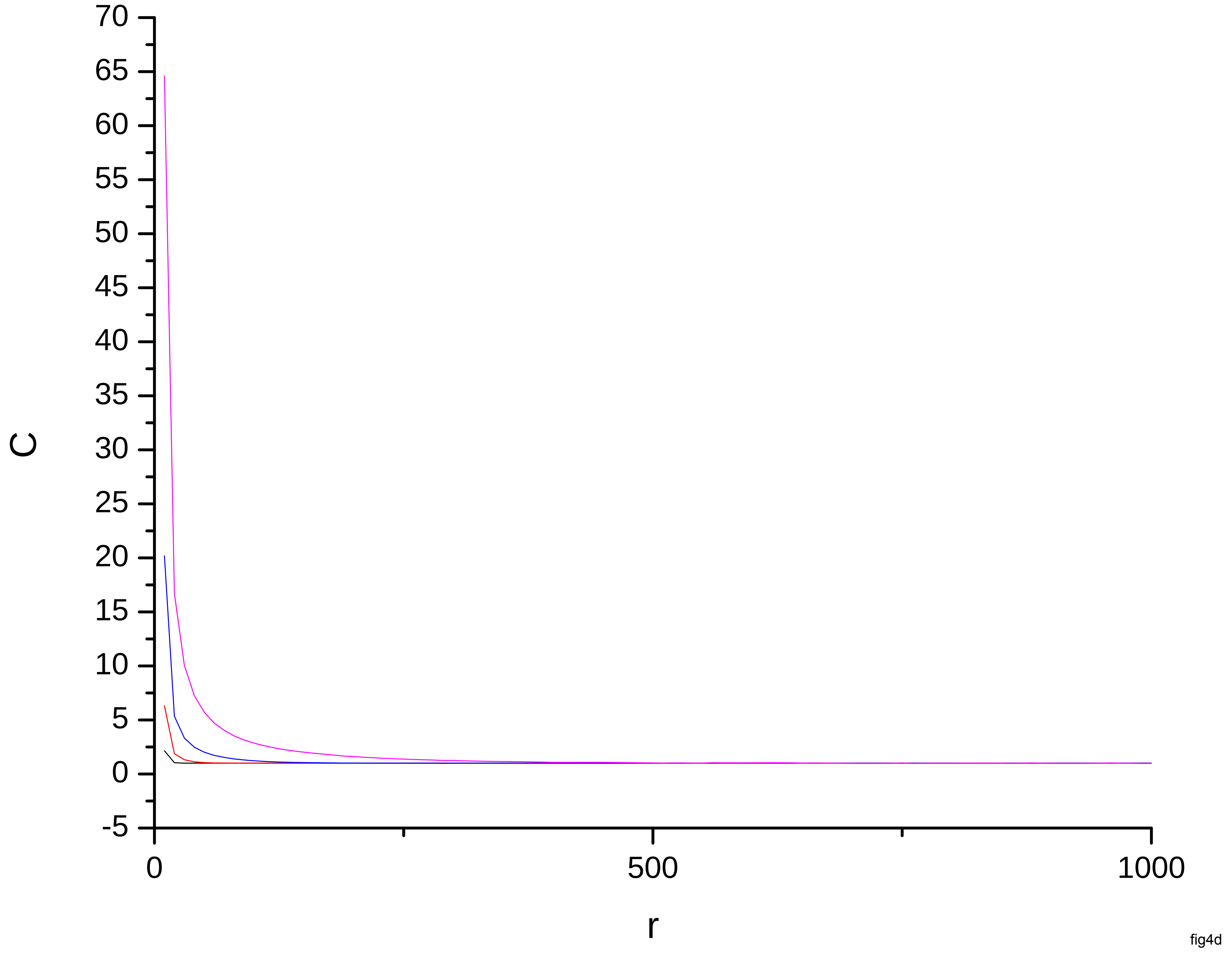

C
70
65
60
55
50
45
40
35
30
25
20
15
10
5
0
-5
0
500
1000
r
fig4d

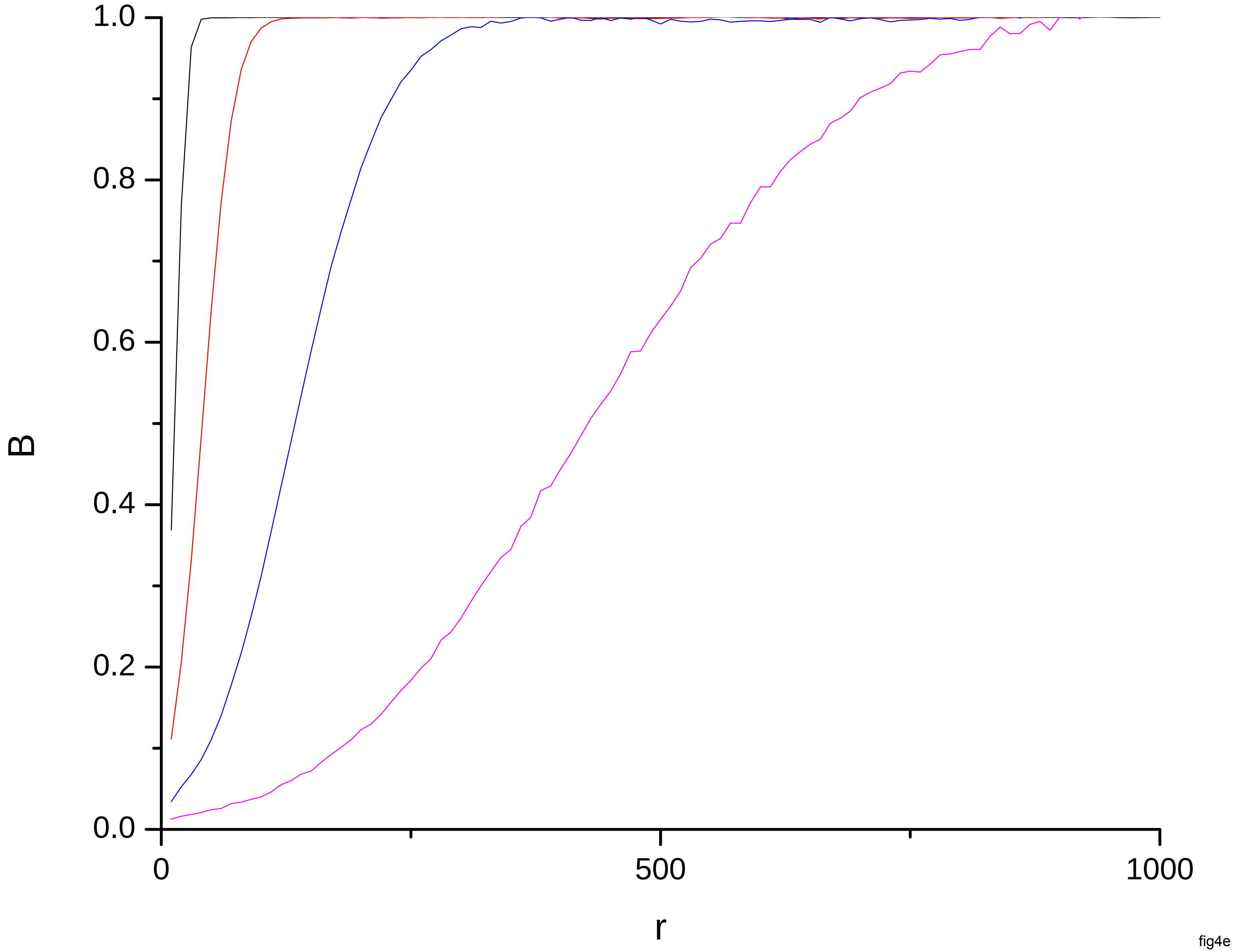

1.0
0.8
0.6
0.4
0.2
0.0
B
0
500
1000
r
fig4e

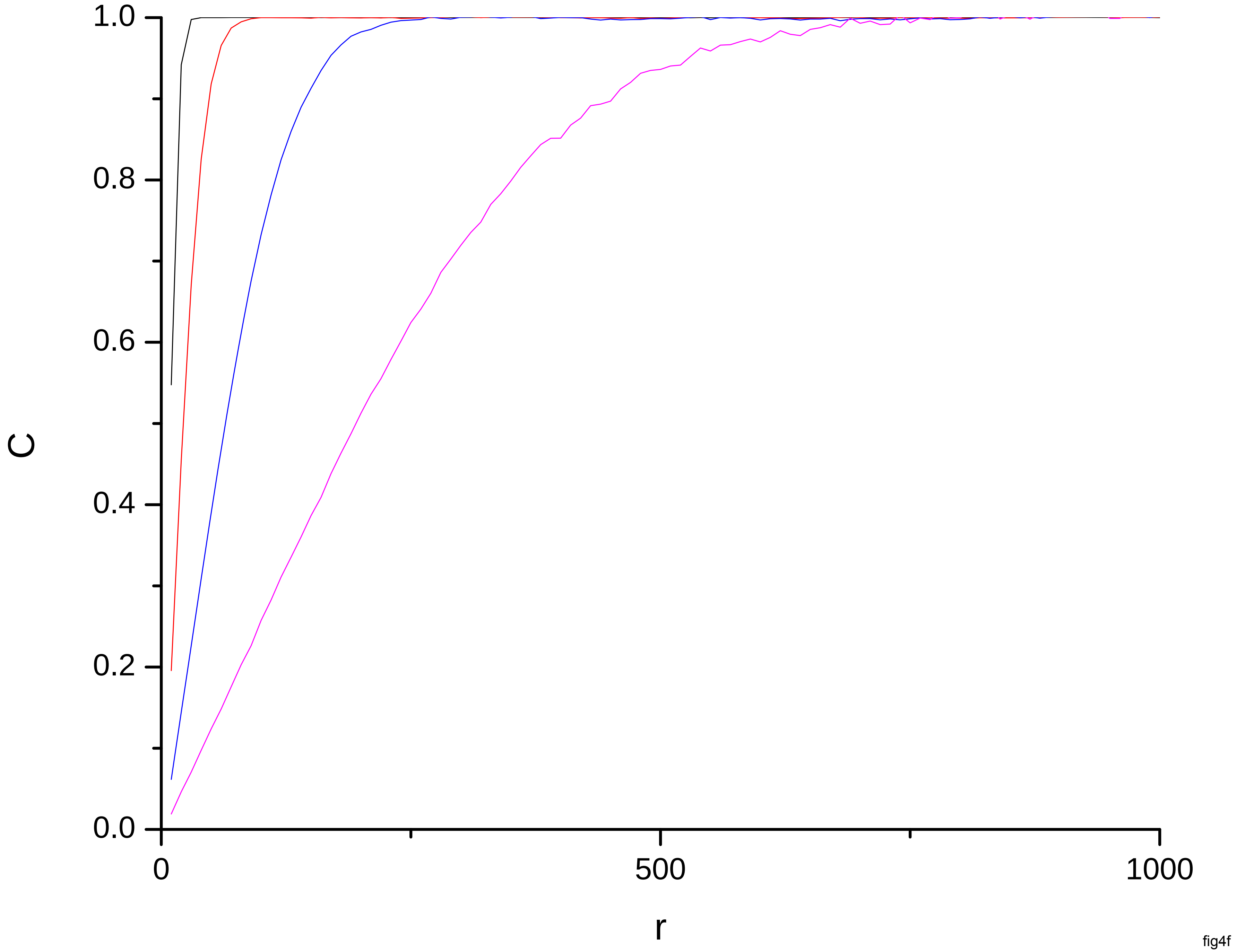

fig4f